# Active Support of Inverters for Improving Short-Term Voltage Security in 100% IBRs-Penetrated Power Systems

Yinhong Lin, *Student Member, IEEE*, Bin Wang, *Senior Member IEEE,* Qinglai Guo, *Fellow, IEEE*, Haotian Zhao*, Member IEEE* and Hongbin Sun*, Fellow, IEEE*

*Abstract*—Due to the energy crisis and environmental pollution, the installed capacity of inverter-based resources (IBRs) in power grids is rapidly increasing, and grid-following control (GFL) is the most prevalent at present. Meanwhile, grid-forming control-based (GFM) devices have been installed in the grid to provide active support for frequency and voltage. In the future GFL devices combined with GFM will be promising, especially in power systems with high penetration or 100% IBRs. When a short-circuit fault occurs in the grid, the controlled current source characteristic of the GFL devices leads to insufficient dynamic voltage support (DVS), while the GFM devices usually reduce the internal voltage to limit the current. Thus, deep voltage sags and undesired disconnections of IBRs may occur. Moreover, due to the dispersed locations and the control strategies' diversity of IBRs, the voltage support of different devices may not be fully coordinated, which is not conducive to short-term voltage security (STVS). To address this issue, a control scheme based on the simulation of transient characteristics of synchronous machines (SMs) is proposed. Then, a new fault ride-through strategy (FRT) is proposed based on the characteristic differences between GFL and GFM devices, and an optimization model of multi-device control parameters is formulated to meet the short-term voltage security constraints (SVSCs) and device capacity constraints. Finally, a fast solution method based on analytical modeling is proposed for the model. Test results based on the double-generator-one-load system, the IEEE 14-bus system, and other systems of different sizes show that the proposed method can effectively enhance the active support capability of GFL and GFM to the grid voltage, and avoid the large-scale disconnection of IBRs.

NOMENCLATURE

*A. Acronyms*

| | |
|---|---|
| DAE | Differential algebraic equation |
| DER | Distributed energy resource |
| DVS | Dynamic voltage support |
| ESS | Energy storage system |
| FRT | Fault ride through |
| GFL | Grid-following control |
| GFM | Grid-forming control |
| IBR | Inverter-based resource |
| NLP | Non-linear programming |
| PCC | Point of common coupling |
| PV | Photovoltaic system |
| SM | Synchronous machine |
| STVS | Short-term voltage security |
| SVSC | Short-term voltage security constraint |

*B. Superscripts and subscripts of variables*

| | |
|---|---|
| $(\cdot)^s$ | Variable under fault s |
| $(\cdot)^0$ | Variable in steady states |
| $(\cdot)_{sp}$ | The given value of a variable |
| $(\cdot)_x / (\cdot)_y / (\cdot)_d / (\cdot)_q$ | x/y/d/q axis component of a variable |
| $(\cdot)_{min} / (\cdot)_{max}$ | Minimum/maximum value of a variable |
| $(\cdot)_m / (\cdot)_l$ | Variables of grid-forming/grid-following control devices |

*C. Sets and parameters*

| | |
|---|---|
| $G / B$ | Real/imaginary part of admittance matrix |
| $\tau_1 / \tau_2 / \tau_3$ | Moment of the fault occurrence/fault steady state/fault clearance |
| $x'_{l,d} / x'_{m,d}$ | Virtual reactance of grid-following and grid-forming control devices |

*D. Variables*

| | |
|---|---|
| $V / I / P / Q$ | Voltage/current/active/reactive power |
| $E / \delta / \theta_{pll} / \theta$ | Internal voltage/power angle/phase angle of PLL/phase angle of voltage |
| $V_{LVRT,th} / V_{HVRT,th}$ | Voltage threshold of PVs |

## I. Introduction

DUE to the energy crisis and environmental pollution, countries around the world are rapidly developing low-carbon energy, especially renewable energy resources represented by wind power and photovoltaic (PV) [1]. For microgrids, the penetration rate of PV is increasing and even reaches 100% in some grids [1], [2]. At the same time, energy storage systems (ESSs) are being installed to curb the impact of renewable energy resources' power fluctuations on power balance, frequency, and voltage regulation [3]. Since many distributed energy resources (DERs) are connected to the grid through inverters, the system is gradually dominated by

This work was supported in part by the Science and Technology Project of State Grid under Grant 5100-202199515A-0-5-ZN.

Y. Lin, B. Wang, Q. Guo, H. Zhao and H. Sun are with the State Key Laboratory of Power Systems, Department of Electrical Engineering, Tsinghua University, Beijing, 100084, China.

inverter-based resources (IBRs) [4]. As synchronous machines (SMs) are being replaced by IBRs gradually, the dynamic voltage support (DVS) capability of the system is becoming insufficient [5]. Under large disturbances such as AC system short-circuit faults, violations of voltage limit at points of common coupling (PCCs) are more likely to occur, causing undesirable disconnection of PV systems [6].

To solve this problem, DVS can be enhanced by installing reactive power compensation devices such as STATCOMs [7]. However, due to the additional cost of installing devices, it is a more economical and promising approach to make use of PV's capability for voltage regulation. Currently, PVs are required to be configured with fault ride-through (FRT) capability in grid codes [8]. Valuable studies have been conducted to improve the DVS capability of PV systems during FRT [9]-[12]. In [9], the DVS capability of PVs is utilized to mitigate short-term voltage instability of the distribution network caused by a large number of transformerless PVs and induction motors. In [10], a DVS control strategy of PVs based on the negative sequence voltage at the PCCs under asymmetrical faults is proposed, which improves the STVS and mitigates the overvoltage extent. In [11], the capacity margins of PVs in the distribution network are utilized to provide more voltage support, thus avoiding IBRs' disconnections due to overcurrent protection. The support effects of STATCOMs and PVs are also compared, and it shows that the proposed scheme can save costs while achieving a similar support effect. In [12], DVS is realized by establishing an additional active and reactive current control mode in the PV controller, which can further improve the STVS compared with the traditional DVS scheme. The above works mainly focus on fault-induced delay voltage recovery (FIDVR) issues [13]. For the issue of PV disconnection caused by instantaneous voltage sags under large disturbances [14], those works may not be applicable. In addition, most of the control strategies of PV systems are grid-following control (GFL) at present [1]. When the penetration rate of renewable energy reaches a certain threshold, the weak grid structures can easily lead to small-signal instability [15], which prevents GFL devices from working properly. Moreover, the insufficient ability of GFL devices to resist voltage and frequency fluctuations [16] is not conducive to system stability.

Due to the above shortcomings, more and more attention has been attracted by the grid-forming control (GFM) strategies [2], [17]-[19], which include virtual synchronous machine control [20], droop control [21], etc. The GFM devices can provide active support to the system voltage and frequency [18], and can maintain stability in weak grids. The application of GFM is an effective way to achieve high penetration of renewable energy in power grids [22]. The MIGRATE projects have shown that 100% penetration of IBRs in power systems can be achieved when more than 30% of devices are applied with GFM [23]. To avoid significant changes in the control structures of conventional PV systems, and to avoid the shortening of the unit's life caused by frequent use of inertia and frequency response, the application of GFM to the ESS converters is more appropriate [24]. It is beneficial to play the role of active and reactive power support from ESSs during the fault period [25]. Therefore, the combined use of GFL and GFM is becoming promising [26], [27]. In this scenario, the utilization of GFL and GFM fast response capability, and the coordination of multiple GFLs and GFMs to provide active voltage support under FRT are essential.

Studies on the application of GFM devices in power grids mainly focus on small-signal stability [28], transient stability [29], frequency stability [30], and so on. There are also a few studies focusing on STVS [2], [20], [31]. In [2], it is shown that the voltage source characteristic of GFM devices is beneficial to improve the STVS of the system. The system voltage recovery performance is improved by improving the control strategy of GFM in [20]. In [31], it is verified that by replacing a SM with a GFM device, the frequency and voltage dynamic characteristics under disturbances can be improved, although the inertia and current tolerance are reduced. To overcome overcurrent issues, various current limiting measures are applied to GFM devices with capacity margins [21], [32]. However, it may cause underutilization of device DVS capability during FRT.

Regarding the co-optimization of multiple IBRs, previous works have focused on inertia and frequency support [27], [33]-[36], FRT [37], voltage control [38], [39], etc. Most of the works focus on the coordination of one type of device, either GFL [27], [36], [37] or GFM devices [34], [35]. A few works have investigated the coordinated control methods for both GFL and GFM devices [27], [33]. These works mainly focus on Volt/Var control at steady states [38], [39]. To the best of our knowledge, there is a gap in the cooperative control of multiple GFLs and GFM devices for improving the STVS of the systems.

In summary, the key issues and challenges still exist when previous works are applied to the scenario where the combination of GFL and GFM devices are utilized for providing active voltage support.

1) IBRs-dominated power systems suffer from weaker DVS and deeper voltage sags. In contrast to synchronous machines (SMs), there is no electromagnetic coupling between windings within the power electronics. Under large disturbances, the dynamic characteristics of the latter are determined by the controllers. Switching the GFL devices to the LVRT control mode under fault disturbances relies on the detection of the voltage amplitudes [40], [41], and a certain time window is required to extract the voltage amplitudes from the instantaneous signal. Therefore, the power responses of the devices lag behind the system voltage changes [42]. Regarding the GFM devices, for protecting power electronics, current limitation strategies are activated under fault disturbances [21], [32]. As a result, IBRs have difficulty in providing sufficient DVS at the instant of a large disturbance. Therefore, the system voltage sags are deeper [43].

2) Various and dispersed IBRs are not fully coordinated, resulting in a degradation of the overall system voltage recovery performance during transient processes. Compared with the traditional power system dominated by a few large-capacity SMs, a distinctive feature of the power system dominated by IBRs is the variety of IBRs' control strategies and the dispersion of IBRs' locations [1]. Since the voltage support effect of different devices may compete with each other [44], massive small-capacity power electronic devices cannot provide the same support effect as a large-capacity unit, and the transient voltage recovery performance of the system is degraded.

3) Co-optimization models of multi-device control parameters tuning considering STVS are difficult to solve efficiently. At various stages of the occurrence of short-circuit faults in the AC system, the system has different demands on the DVS of the devices. Therefore, it is necessary to consider the characteristics of different stages of the transient processes for parameter tuning. However, the original mathematical model describing the complete dynamic processes is a set of differential-algebraic equations (DAEs). Moreover, due to the larger number of IBRs compared to SMs, the original mathematical model contains higher dimensional DAEs. The solution processes will suffer from computational efficiency issues in real applications [45].

The main ideas and contributions of this paper are as follows.

**1) A DVS enhancement control strategy based on the simulation of transient characteristics of SMs is proposed.** The transient mathematical model of the SM is formulated in the reference coordinate system of the GFL and GFM devices, respectively. It is compared with the conventional GFL and GFM controllers and the "missing capability" is identified. The inner control loop of the power electronics devices is modified. With the help of fast inner-loop control, the deficiency of DVS of power electronic devices is compensated. At the same time, by appropriately tuning the control parameters, stronger DVS can be tapped within the device capacity.

**2) A new FRT strategy as well as an optimization model for tuning of control parameters of multi-GFL and GFM devices considering short-term voltage security constraints (SVSCs) is proposed.** Based on the controlled voltage source and current source characteristics of GFM and GFL devices under large disturbances [46], [47], the output voltage and current reference values during fault periods are tuned in advance. By establishing a multi-device control parameter optimization model that considers the SVSCs and device capacity constraints, the control parameters of different devices are tuned in advance before the fault occurrences. When a large disturbance occurs, the devices enter the FRT mode and track the pre-set values, and coordination between different devices is achieved.

**3) An analytical modeling approach based on the dynamic characteristics of devices in multiple time scales is proposed.** Transient voltage violation mainly occurs at fault occurrences, fault steady states, and fault clearances. Therefore, these critical moments are formulated separately. According to the singular perturbation theory [48], dynamics processes that are much faster than the power frequency are completed instantaneously, and the corresponding DAEs degenerate into algebraic equations. For much slower processes, it is assumed that they have not yet occurred and the corresponding state variables remain constant. Therefore, the optimization model is simplified into a general NLP problem. Through equivalent transformations, the Hessian matrix of the model is constant, and the model can be solved efficiently.

The remainder of this paper is organized as follows. Section II proposes the enhancement control strategy for GFL and GFM devices. Section III proposes a coordinated FRT strategy for IBRs. Section IV proposes a fast solution method for the model. Numerical simulations are conducted in Section V. Section VI summarizes the conclusions.

## II. ENHANCED CONTROL METHOD FOR IMPROVING DYNAMIC VOLTAGE SUPPORT CAPABILITY

### A. Analysis of transient characteristics and modeling

From the power frequency time scale, the instantaneous characteristics of a SM under a short-circuit fault are determined by the electromagnetic coupling between the stator and rotor windings [49]. In contrast, for GFL devices, the transient characteristics under short-circuit faults are determined by fast current inner loops. The outer-loop control mode switching relies on the detection of the three-phase voltage amplitudes [40], and the detection requires a certain time window. As a result, the GFL devices' injection current lags behind the changes of voltages at PCCs, and the GFLs track the current reference value generated before the occurrence of large disturbances, thus the transient support capability is weak.

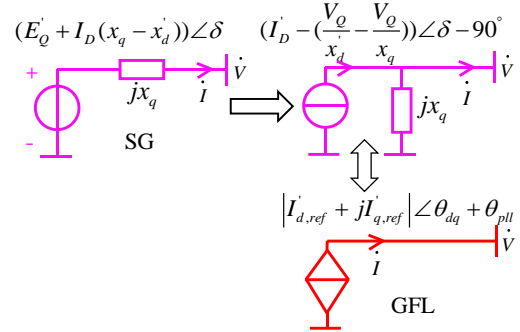

Fig. 1. Transient models of a SM and GFL device under Large Disturbances.

To compare the differences in the large disturbance characteristics of the two devices, the transient model of the SM (1) is given in [49] and transformed into a Norton equivalent circuit (2), while the controlled current source of the GFL is also presented, as shown in Fig. 1. The equivalent current $I'_D$ is maintained constant when a large disturbance occurs, which is like the property of the GFL device that the injected current $|I'_{d,ref} + jI'_{q,ref}|$ is constant at the instant of large disturbances. However, there are also some differences between the models of SMs and power electronics. $V_Q(1/x'_d - 1/x_q)$ characterizes the effect of electromagnetic induction, and $\dot{V}/jx_q$ characterizes the increase in current due to the voltage sag. When a short-circuit fault occurs, the injection current of a SM will change instantaneously due to the instantaneous change of the voltage $V_Q$ and $V_D$ at the PCCs. In contrast, for GFL devices, the lack of these terms results in a weak spontaneous response.

$$\dot{V} = (E'_Q + I_D(x_q - x'_d))\angle\delta - j\dot{I}x_q \\
= \dot{E}'_Q - jx'_d \dot{I}_D - jx_q \dot{I}_Q \quad (1)$$

$$\begin{aligned}\dot{I} &= (E'_Q / x'_d - V_Q(1/x'_d - 1/x_q))\angle(\delta - 90°) - \dot{V}/jx_q \\ &= \dot{I}'_D - \dot{V}_Q/jx'_d - \dot{V}_D/jx_q\end{aligned} \quad (2)$$

### B. Enhanced control strategy

To simulate the effect of the spontaneous response of the SM as much as possible, the SM transient model (2) is reformulated in the inverter dq-axis coordinate system, as is shown in Fig. 2 and Eq. (3).

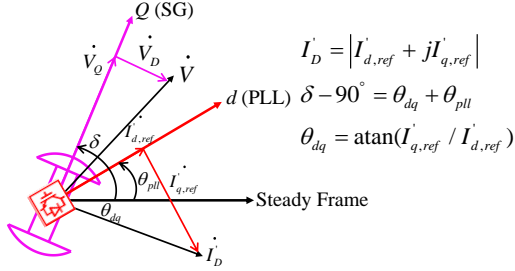

Fig. 2. SMs' model in the reference coordinate system of GFL devices.

Note that the goal of this paper is to enhance the DVS capability of the inverter rather than to fully simulate the dynamic characteristics of the SM. Therefore, to simplify the inverter controllers, by substituting $x'_d = x_q$ into (3), the enhanced control strategy is shown in Fig. 3, which is equivalent to a parallel "virtual reactance" at the terminal. When the device capacity is large enough, from the time scale of the power frequency, the synthesized current references increase rapidly due to the rapid decrease of the terminal voltage under large disturbances. As a result, the device can exhibit a stronger DVS capability. However, limited by the device capacity, the control parameters need to be optimized to prevent the injection current from saturating.

$$\begin{aligned} I_{d,ref} &= I'_{d,ref} - \sin\theta_{dq}\cos\theta_{dq}(1/x_q - 1/x'_d)V_d \\ &\quad - (\cos^2\theta_{dq}/x'_d + \sin^2\theta_{dq}/x_q)V_q \\ I_{q,ref} &= I'_{q,ref} + (\cos^2\theta_{dq}/x_q + \sin^2\theta_{dq}/x'_d)V_d \\ &\quad + \sin\theta_{dq}\cos\theta_{dq}(1/x_q - 1/x'_d)V_q \end{aligned} \quad (3)$$

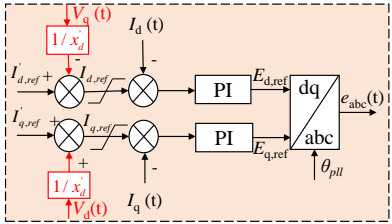

Fig. 3. Improved inner loop of GFL devices.

Similar to the above process, the SM model is also reformulated in the GFM reference coordinate system. By substituting $E'_Q = E_{ref}$, $\delta_g = \delta_{GFM}$, $x'_d = x_q$, the improved controller is shown in Fig. 4. The physical meaning is similar to the "virtual reactance", which can avoid overcurrent at the instant of fault occurrence through the voltage drop on the virtual reactance [32]. The difference is that in this paper, the control parameters $x'_d$ will be optimized considering anticipated contingencies, and the dynamic performance will be improved as much as possible within the current limit.

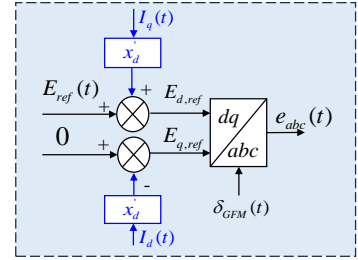

Fig. 4. Improved inner loop of GFM devices.

## III. MULTI-GFL AND GFM DEVICES FAULT RIDE THROUGH COOPERATIVE CONTROL METHOD

### A. Control strategy and switching logic

GFL devices exhibit controlled current source characteristics under large disturbances, while GFM devices exhibit controlled voltage source characteristics. To fully coordinate the DVS capability of these devices, the optimal current and voltage reference values can be set in advance before the fault occurrence. When the fault occurs, the controllers switch to the proposed FRT mode when some conditions are met, as shown in Fig. 5. Where mode 1 is the normal mode, mode 2 is the proposed FRT mode, and mode 3 is the transition mode. The specific process is analyzed as follows.

When a fault occurs, the GFL and GFM devices will respond to disturbances quickly through the inner loop. It can be regarded as the transient response of SMs. The main parameters affecting the dynamic characteristics at this moment are $(x'_{d,GFL}, x'_{d,GFM})$. Then, the outer loop of the controller detects that the terminal voltage is below a certain threshold and will switch to the proposed FRT mode. At this moment, both GFL and GFM devices' outer loop modules are frozen. They track the pre-set reference generated before the accident. Then, when the fault is cleared, the system voltage is restored, and the injected current and output voltage amplitude of the devices transition exponentially to the pre-freezing reference. The phase angle and frequency of the GFM devices track the voltage phase angle and frequency of PCCs, respectively [50]. Finally, when they are close to each other, the devices switch back to the normal control mode.

The proposed control framework is shown in Fig. 6. Three parameters need to be optimized for each inverter. Those parameters have a great impact on the voltage sag at the instant of fault occurrences, the voltage level at fault steady states, and the overvoltage at the instant of fault clearances. Hence, an optimization model for co-optimizing different inverter control parameters will be formulated.

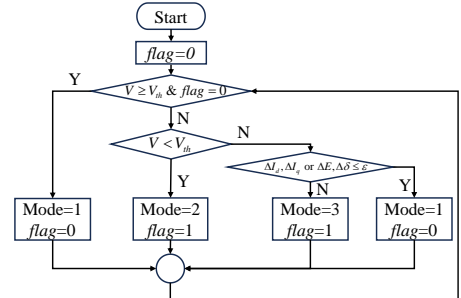

Fig. 5. Control mode switching logic.

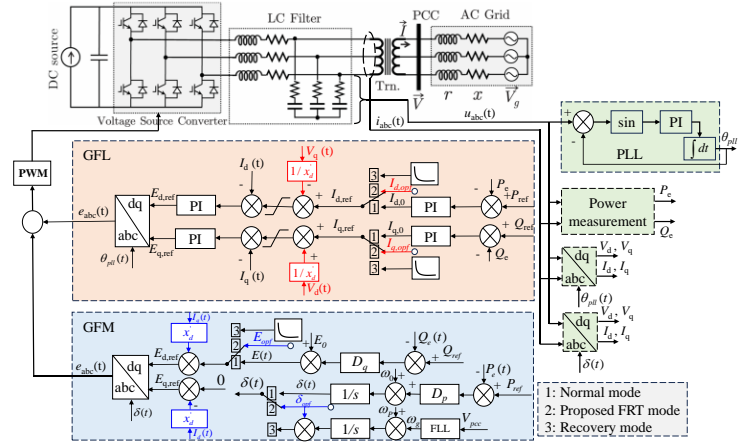

Fig. 6. Overall framework of GFL and GFM device controllers

### B. Original optimization model

Both GFL and GFM devices contain several parameters to be optimized, which have a significant impact on the dynamic characteristics of the devices during the fault process. For tuning these parameters, it is necessary to consider the capacity limit of the power electronics and the SVSCs. In this case, the optimization objective is to improve the voltage recovery performance of the weak points of the systems. The compact form of the original model is shown in (5)-(10).

$$\min S(\boldsymbol{u}) \tag{4}$$

$$\text{s.t. } \boldsymbol{G}^{t,0}\left(\boldsymbol{u}^t, \boldsymbol{x}^{t,0}, \boldsymbol{y}^{t,0}\right) = \boldsymbol{0} \tag{5}$$

$$\underline{\boldsymbol{H}_1} \leqslant \boldsymbol{H}_1\left(\boldsymbol{u}^t, \boldsymbol{x}^{t,0}, \boldsymbol{y}^{t,0}\right) \leqslant \overline{\boldsymbol{H}_1} \tag{6}$$

$$d\boldsymbol{x}^{t,s}(\tau)/d\tau = f^{t,s}(\boldsymbol{u}^t, \boldsymbol{x}^{t,s}(\tau), \boldsymbol{y}^{t,s}(\tau)) \tag{7}$$

$$\boldsymbol{g}^{t,s}(\boldsymbol{x}^{t,s}(\tau), \boldsymbol{y}^{t,s}(\tau), \boldsymbol{u}^t) = \boldsymbol{0} \tag{8}$$

$$\underline{\boldsymbol{H}_2}(\tau) \leqslant \boldsymbol{H}_2\left(\boldsymbol{x}^{t,s}(\tau), \boldsymbol{y}^{t,s}(\tau)\right) \leqslant \overline{\boldsymbol{H}_2}(\tau) \tag{9}$$

$$\tau \in [\tau_1, \tau_2], \forall t \in S_T, \forall s \in S_{flt}$$

(5) denotes the optimization objective is to minimize the deviation of the weak point voltage from the reference values. (6) denotes the steady-state power flow. (7) denotes the steady-state security constraints as well as the control variable ranges. (8) and (9) denote the set of DAEs describing the dynamic characteristics of the system. In normal mode, for GFL devices, the outer-loop PI control and inner-loop PI control are applied [51]. For GFM devices, the frequency and voltage droop control is applied [52]. In the proposed FRT mode, the outer loops of the inverters are frozen and the inner loops track the pre-set values. In the recovery mode, the output voltage and current of the inverters transition from the optimized references to the pre-freezing references exponentially. Meanwhile, the output frequency and phase angle of GFM devices tracks the frequency and voltage phase angle at PCCs, respectively [50].

## IV. EFFICIENT SOLUTION METHOD BASED ON ANALYTICAL MODELING APPROACH

The above model containing DAEs is difficult to solve directly. Some previous works remove the DAEs from the model and solve it by time-domain simulations. Then SVSCs are modeled based on the trajectory sensitivity method [53]. Due to time-consuming electromagnetic transient simulations, the trajectory sensitivity method faces the challenge of low computational efficiency. Therefore, a practical solution method is necessary for practical applications. Based on a large number of time-domain simulations, it can be found that the voltage violations are likely to occur at the instants of fault occurrences, the fault steady states, and the instants of fault clearances. Therefore, for simplification, analytical modeling of these critical moments is carried out in the following.

The idea of analytical modeling is inspired by the singular perturbation theory [48]. For a system containing multi-timescale dynamics, the fast-scale dynamic process is completed instantaneously if it is much faster compared to the time scale of interest. The corresponding DAEs degenerate into algebraic equations. If the other dynamical process is much slower, the corresponding state variables remain constant.

From the time scale of power frequency, at the instant of fault occurrences, since the control mode switching depends on the voltage amplitude detection and requires a time window, the inner loop still tracks the reference generated before the fault. If the output current of the device does not exceed the capacity limit, the response characteristics of the GFM and GFL devices at the instant of fault occurrences can be considered as a controlled voltage source and a controlled current source, respectively, which can be described by algebraic equations. At the later stage of the fault period, since the duration of the fault is usually hundreds of milliseconds, in the proposed FRT mode, the GFM and the GFL devices have already tracked the preset references and reached a new steady state, which can be described by algebraic equations. At the instant of fault clearances, due to the delay in voltage detection, the controller is still in the proposed FRT mode, and the output current and voltage of the device are still tracking the preset references, which can be described by algebraic equations. Based on the above analysis, the following analytical model can be obtained. The objective function is to minimize the deviation of the weak node voltages from the voltage references at the three critical moments (11).

*1) Objective function*

$$\min_{x_{l,d}^{\cdot}, x_{m,d}^{\cdot}, E_{m,opf}, \delta_{m,opf}, I_{l,d,opf}, I_{l,q,opf}} \sum_{\tau} \left| V_{ref} - V(\tau) \right| \tag{10}$$

s.t. (11)-(23)

The equality constraints and inequality constraints corresponding to GFL and GFM devices as well as the power grids are as follows. Where the initial values of operating variables are given and are not involved in the optimization. Only the control parameters related to the dynamic processes are optimized.

*2) Voltage equation for GFM devices in steady state*
$$V_{m,x,sp}^0 = E_{m,x}^{'0} + I_{m,y,sp}^0 x_{m,d}^{'}, V_{m,y,sp}^0 = E_{m,y}^{'0} - I_{m,x,sp}^0 x_{m,d}^{'} \quad (11)$$

*3) Current equation for GFL devices in steady state*
$$I_{l,x,sp}^0 = I_{l,x}^{'0} - V_{l,y,sp}^0 / x_{l,d}^{'}, I_{l,y,sp}^0 = I_{l,y}^{'0} + V_{l,x,sp}^0 / x_{l,d}^{'} \quad (12)$$

*4) Coordinate transformation equations in steady state*
$$T_m^0 = \begin{bmatrix} \sin\delta_m^0 & -\cos\delta_m^0 \\ \cos\delta_m^0 & \sin\delta_m^0 \end{bmatrix}, T_l^0 = \begin{bmatrix} \sin\theta_{l,pll}^0 & -\cos\theta_{l,pll}^0 \\ \cos\theta_{l,pll}^0 & \sin\theta_{l,pll}^0 \end{bmatrix} \quad (13)$$

$$\boldsymbol{E}_{m,dq}^{'0} = T_m^0 \boldsymbol{E}_{m,xy}^{'0}, \boldsymbol{I}_{l,dq}^{'0} = T_l^0 \boldsymbol{I}_{l,xy}^{'0} (\forall m \in S_{GFM}, \forall l \in S_{GFL})$$

*5) Voltage equation of GFM devices during transient process*
$$V_{m,x}^s(\tau) = E_{m,x}^{'s}(\tau) + I_{m,y}^s(\tau)x_{m,d}^{'}, V_{m,y}^s(\tau) = E_{m,y}^{'s}(\tau) - I_{m,x}^s(\tau)x_{m,d}^{'} \quad (14)$$

*6) Current equation of GFL devices during transient process*
$$I_{l,x}^s(\tau) = I_{l,x}^{'s}(\tau) - V_{l,y}^s(\tau)x_{l,d}^{'}, I_{l,y}^s(\tau) = I_{l,y}^{'s}(\tau) + V_{l,y}^s(\tau)x_{l,d}^{'} \quad (15)$$

*7) Coordinate transformation equations in transient process*
$$T_m^s(\tau) = \begin{bmatrix} \sin\delta_m^s(\tau) & -\cos\delta_m^s(\tau) \\ \cos\delta_m^s(\tau) & \sin\delta_m^s(\tau) \end{bmatrix}$$
$$T_l^s(\tau) = \begin{bmatrix} \sin\theta_{l,pll}^s(\tau) & -\cos\theta_{l,pll}^s(\tau) \\ \cos\theta_{l,pll}^s(\tau) & \sin\theta_{l,pll}^s(\tau) \end{bmatrix} \quad (16)$$

$$\boldsymbol{E}_{m,dq}^{'s}(\tau) = T_m^s(\tau)\boldsymbol{E}_{m,xy}^{'s}(\tau), \boldsymbol{I}_{l,dq}^{'s}(\tau) = T_l^s(\tau)\boldsymbol{I}_{l,xy}^{'s}(\tau)$$

*8) Network equations in transient process*
$$\begin{bmatrix} I_x^s(\tau) \\ I_y^s(\tau) \end{bmatrix} = \begin{bmatrix} G^s(\tau) & -B^s(\tau) \\ B^s(\tau) & G^s(\tau) \end{bmatrix} \begin{bmatrix} V_x^s(\tau) \\ V_y^s(\tau) \end{bmatrix} \quad (17)$$

$$V_x^{s2}(\tau) + V_y^{s2}(\tau) = V^{s2}(\tau) \quad (18)$$

$$\tan\theta^s(\tau) = V_y^s(\tau)/V_x^s(\tau) \quad (19)$$

*9) Security constraints of IBRs*
$$I_{l,x}^{s2}(\tau) + I_{l,y}^{s2}(\tau) \leq I_{l,\max}^2, I_{m,x}^{s2}(\tau) + I_{m,y}^{s2}(\tau) \leq I_{m,\max}^2 \quad (20)$$

$$V_{LVRT,th} \leq V_l^s(\tau) \leq V_{HVRT,th}, V_{LVRT,th} \leq V_m^s(\tau) \leq V_{HVRT,th} \quad (21)$$

*10) Boundary conditions of GFM devices*
$$E_m^{'s}(\tau_1) = E_m^{'0}, \delta_m^s(\tau_1) = \delta_m^0$$
$$E_m^{'s}(\tau_3) = E_m^{'s}(\tau_2) = E_{m,opf}^{'s}, \delta_m^s(\tau_3) = \delta_m^s(\tau_2) = \delta_{m,opf}^s \quad (22)$$

*11) Boundary conditions of GFL devices*
$$I_{l,d}^{'s}(\tau_1) = I_{l,d}^{'0}, I_{l,q}^{'s}(\tau_1) = I_{l,q}^{'0}, \theta_{l,pll}^s(\tau_1) = \theta_{l,pll}^0 = \theta_{l,sp}^0$$
$$I_{l,d}^{'s}(\tau_3) = I_{l,d}^{'s}(\tau_2) = I_{l,d,opf}^{'s}, I_{l,q}^{'s}(\tau_3) = I_{l,q}^{'s}(\tau_2) = I_{l,q,opf}^{'s} \quad (23)$$
$$\theta_{l,pll}^s(\tau_3) = \theta_{l,pll}^s(\tau_2) = \theta_l^s(\tau_2)$$

Note that although the model is nonconvex, the rest of the model is quadratic except for the voltage phase angle constraint (20) and the coordinate transformation equations (14)(17). Since the power angle and phase-locked loop angle cannot change suddenly at the moment of large disturbances, the dq-axis current reference and voltage reference given by the outer loop cannot be changed suddenly. Therefore, the xy-axis current and voltage reference cannot change suddenly as well. To solve the model more efficiently, the xy-axis components of the injected currents and voltages can be regarded as optimization variables, and (14)(17)(20) are removed. Meanwhile, the boundary conditions (23)(24) at different critical moments are replaced by (25)(26). Since the Hessian matrix of the model is a constant matrix, it can be solved efficiently by the interior point method [54].

$$E_{m,x}^{'s}(\tau_1) = E_{m,x}^{'0}, E_{m,y}^{'s}(\tau_1) = E_{m,y}^{'0}, E_{m,x}^{'s}(\tau_3) = E_{m,x}^{'s}(\tau_2) = E_{m,x,opf}^{'s}$$
$$E_{m,y}^{'s}(\tau_3) = E_{m,y}^{'s}(\tau_2) = E_{m,y,opf}^{'s}$$
(24)

$$I_{l,x}^{'s}(\tau_1) = I_{l,x}^{'0}, I_{l,y}^{'s}(\tau_1) = I_{l,y}^{'0}, I_{l,x}^{'s}(\tau_3) = I_{l,x}^{'s}(\tau_2) = I_{l,x,opf}^{'s} \quad (25)$$
$$I_{l,y}^{'s}(\tau_3) = I_{l,y}^{'s}(\tau_2) = I_{l,y,opf}^{'s}$$

## V. CASE STUDY

In this section, numerical simulations are conducted on MATLAB/SIMULINK to validate the proposed method. The simulations are conducted using a LEGION laptop with an Intel i7-9750H CPU and 16 GB of RAM.

### A. Performance of the enhanced control strategies

In this subsection, a two-generator-one-load system, as shown in Fig. 7, is utilized to validate the effectiveness of the proposed enhanced control strategy. For comparison, in the first case, the GFM devices are controlled by the common droop control, and the GFL devices are controlled by the common FRT control [32]. In the second case, the GFL and GFM devices are controlled by the proposed control strategy. The comparison results are shown in Fig. 8-Fig. 9.

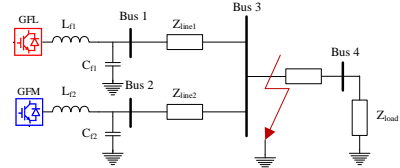

Fig. 7. Schematic diagram of a double-generator-one-load-system.

The dynamic responses of the terminal voltage of the GFM, the output current of the GFL, the outer-loop input signals of the GFL and the GFM, and the inner-loop input signal of the GFL are presented in Figs. 8(a)-(d). From Fig. 8(a), the GFM output voltage amplitude undergoes a transient sag at the instant of fault occurrence, thus avoiding the overcurrent. This is caused by an increase in the voltage drop across the virtual reactance. However, at this moment the voltage at the PCC drops to 0.18 p.u. and the device is at risk of disconnection. In contrast, from Fig. 9(b), at the instant of the fault, the device injected current is unable to respond instantaneously with the voltage drop at the terminal, and there is an upward process before it finally reaches a peak value of 1.2 p.u. From Fig. 8(c), neither the power generation nor the voltage magnitude extracted from the instantaneous signal jumps at the instant of fault. As shown in Fig. 8(d), the injected current in Fig. 8(b) is maintained at the pre-accident level due to the unchanged input signal of the inner loop at the fault instant. Then, the GFL device enters the LVRT control mode to increase the reactive injected current while decreasing the active current to avoid overcurrent. As a result, the effect of the device in supporting the system voltage at the instant of fault is not strong, and the voltage drop is large.

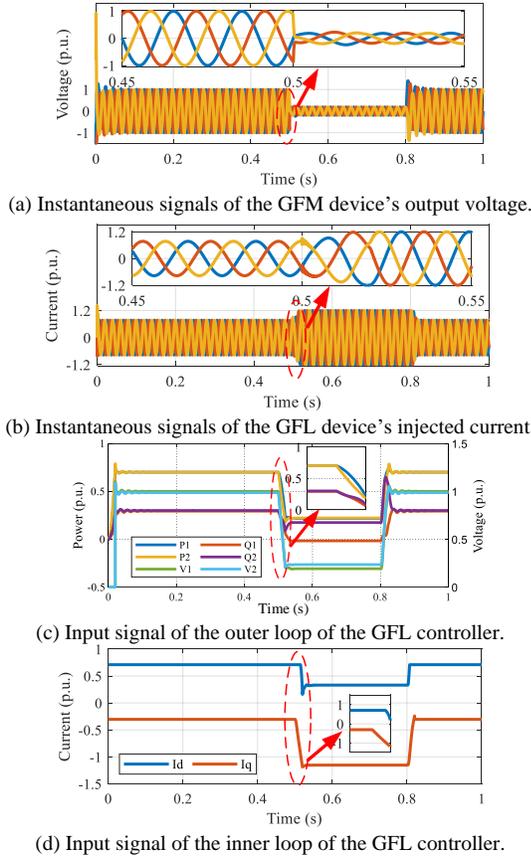

(a) Instantaneous signals of the GFM device's output voltage.

(b) Instantaneous signals of the GFL device's injected current.

(c) Input signal of the outer loop of the GFL controller.

(d) Input signal of the inner loop of the GFL controller.

Fig. 8. Dynamic responses of the IBRs under common FRT strategy.

The dynamic responses of the output voltage of the GFM, the output current of the GFL, the outer-loop input signals of the GFL and the GFM, and the inner-loop input signal of the GFL are shown in Figs. 9(a)-(d). From Fig. 9(a), the transient drop in the terminal voltage occurs at the instant of fault, but the voltage drop is smaller and elevated to 0.23 p.u. due to the stronger DVS of the GFL devices at this time. From Fig. 9(b), the injected current of the GFL increases rapidly to its peak value, i.e., 1.2 p.u., at the instant of the fault. Therefore, the device can provide very fast voltage support to the grid. From Fig. 9(c), the voltage magnitude as well as the active and reactive signals extracted by the controllers from the instantaneous signals do not change. From Fig. 9(d), the current reference signal of the outer loop changes slowly at the instant of fault. However, the inner loop synthesized reference signal shows a step due to the correction term. Therefore, the device can change the injected current by the fast response of the current inner loop under large disturbances to increase the injected current and resist the deep voltage drop, thus preventing the IBR from disconnection.

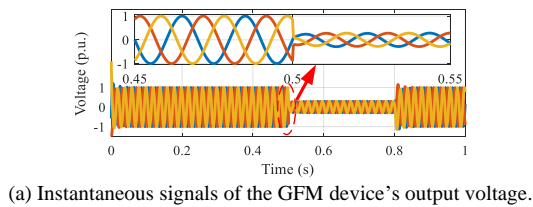

(a) Instantaneous signals of the GFM device's output voltage.

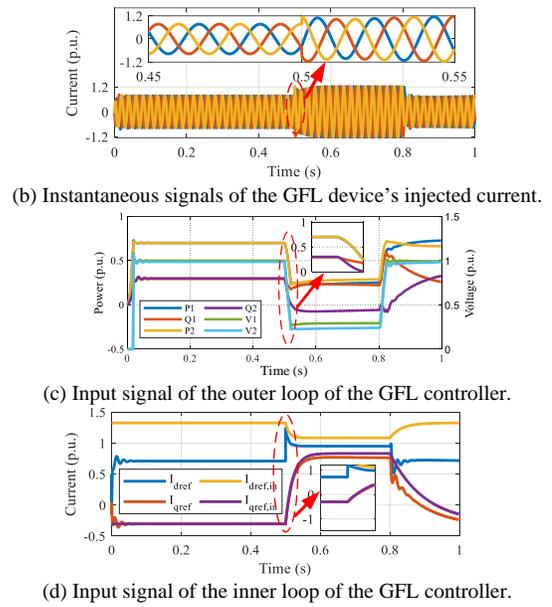

(b) Instantaneous signals of the GFL device's injected current.

(c) Input signal of the outer loop of the GFL controller.

(d) Input signal of the inner loop of the GFL controller.

Fig. 9. Dynamic response of the IBRs with proposed FRT strategy.

Fig. 10(a) and (b) present the voltage amplitude response when different control parameters are adopted for the two control strategies, respectively. From Fig. 10(a), under FRT, the degree of terminal voltage drop at the instant of fault remains unchanged even if different reactive current gains are adopted. This is because the device does not enter the FRT mode when the voltage drops suddenly. In contrast, it can be seen from Fig. 10(b) that at the instant of fault, the depth of terminal voltage sag decreases as the virtual reactance decreases, which indicates that the DVS capability is enhanced. However, a smaller virtual reactance may lead to current saturation and it affects the voltage swell level at the moment of fault clearances. Therefore, the control parameters need to be optimized considering the device capacity as well as the voltage security at different stages of the transient process.

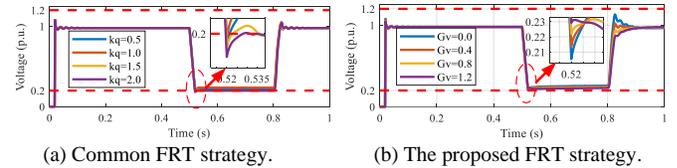

(a) Common FRT strategy.   (b) The proposed FRT strategy.

Fig. 10. Dynamic responses of the inverter with different control parameters.

*B. Effectiveness of the proposed methodology*

To verify the feasibility of the optimization method proposed in this paper, a modified IEEE 14-bus system is used [55]. In this case, bus 2 and 8 are installed with two GFL devices, and bus 1, 3, and 6 are installed with three GFM devices. The terminal voltage response of different IBRs, the phasor decomposition of the weak node voltage, the output voltage and injected current response of the GFM devices, and the injected active and reactive current response of the GFL devices are presented in Figs. 11 and 12, respectively.

From Fig. 11(a), when a common FRT strategy is applied to GFL and GFM devices, at the instant of fault occurrence, the devices GFM1, GFL2, and GFM3 terminal voltages drop to a greater extent. Among them, GFL2 terminal voltage drops to 0.179 p.u., and GFL2 has a greater risk of disconnection. Moreover, the terminal voltage remained at a low level of 0.182

p.u. during the fault period. The terminal voltage of some units is close to 1.2 p.u. at the instant of fault clearances, and there is a transient overvoltage issue. As shown in Fig. 11(b), the voltage support effect from different IBRs compete with each other, and the voltage on bus 5 is only 0.16 p.u. This is because the common strategy reacts to the fault disturbance through the local response of the IBRs, which does not take into account the differences in the locations and control strategies of different IBRs. From Fig. 11(c), the potential of the GFM with conventional strategy remains constant during the fault period to avoid overcurrent. However, its injection current is only 0.615 p.u., indicating that its support capability can be further improved. From Fig. 11(d), the GFL reactive current cannot grow rapidly at the instant of fault, which is caused by the delay in the voltage amplitude detection. Moreover, there is a delay in the reactive current retraction at the instant of fault clearances, which leads to the overvoltage issue.

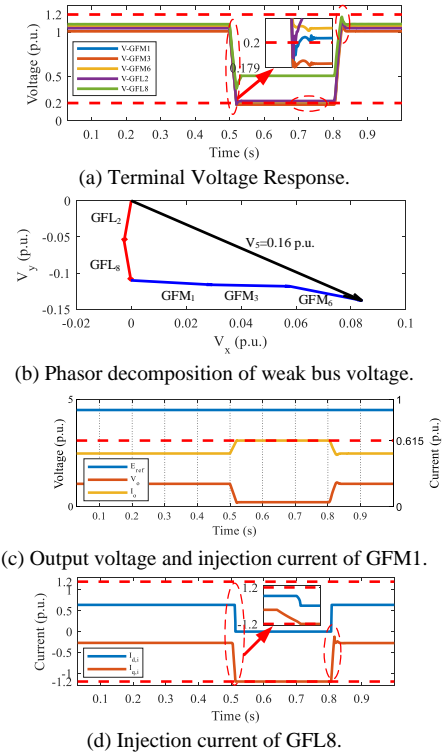

(a) Terminal Voltage Response.

(b) Phasor decomposition of weak bus voltage.

(c) Output voltage and injection current of GFM1.

(d) Injection current of GFL8.

Fig. 11. Dynamic response under FRT control method

From Fig. 12(a), the overall voltage levels of the system are improved at the instant of fault occurrence through the proposed method. The terminal voltages of the IBRs are above 0.23 p.u., which avoids the IBRs' disconnection. Moreover, the voltage level is improved during the fault steady state, and the overvoltage issue is effectively mitigated when the fault is cleared. From Fig. 12(b), during the fault steady state, the voltage support effect of different IBRs is coordinated by the proposed method, and the voltages have been improved to 0.28 p.u. From Fig. 12(c), the injected current of the GFM device at the instant of fault is rapidly increased to the capacity limit, i.e., 1.2 p.u., which effectively improves the system voltage level at the instant of fault. From Fig. 12(d), after the fault occurs, the injected current of the GFL device changes rapidly, which resists the terminal voltage drop. Moreover, the current does not exceed the device capacity limit. During the fault steady state, the injected active and reactive currents of the GFL device reach a pre-set value, allowing the system voltage to be maintained at a higher level. This indicates that the proposed method can fully coordinate the active and reactive power of different GFL and GFM devices.

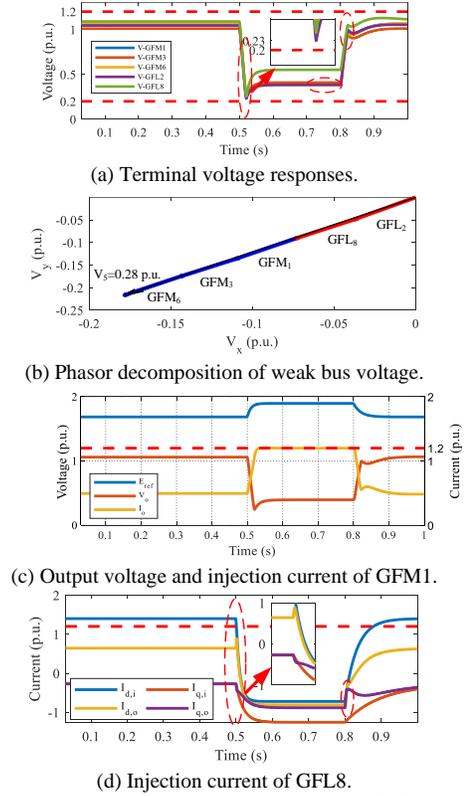

(a) Terminal voltage responses.

(b) Phasor decomposition of weak bus voltage.

(c) Output voltage and injection current of GFM1.

(d) Injection current of GFL8.

Fig. 12. Dynamic responses under the proposed FRT method.

*C. Accuracy of analytical modeling*

Based on the modified IEEE 14-bus system, the accuracy of the analytical model for voltage magnitudes during the transient process is verified. All bus voltage magnitudes are calculated by the proposed method at the critical moments under different faults. Meanwhile, accurate electromagnetic transient simulations are conducted for comparative analysis, and the errors are shown in Fig. 13(a). Where the light blue shading indicates the range of errors and the dark blue dots indicate the average errors. It can be found that for the voltage amplitudes at the three critical moments, the maximum absolute errors of the proposed method are less than 4e-03p.u., 4e-04p.u., and 3.5e-03p.u. under different faults, respectively, which are within an acceptable range. The simulation results of the voltage responses under a short-circuit fault at bus 4 and the calculation results of the analytical model are shown in Fig. 13(b). It can be found that the results of the analytical model nearly coincide with the time-domain simulation results, which indicates that the proposed analytical model has high accuracy. Therefore, the analytical model can be used to establish SVSCs at some critical moments during the fault process, thus avoiding the execution of a large number of time-domain simulations in industrial applications.

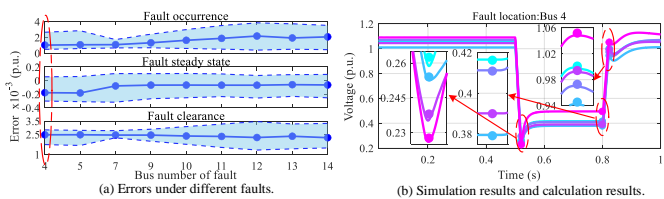
Fig. 13. Errors analysis under different faults.

*D. Scalability Analysis*

The model proposed in this paper does not contain integer variables, it is a general NLP model. The nonlinear terms are the bilinear terms as well as the squared terms. Therefore, its Hessian matrix is a constant coefficient matrix, which can be solved efficiently using the interior point method. With the help of state-of-the-art solvers like IPOPT [54], systems of different sizes are tested and their solution results are shown in Tab. 1.

TABLE I COMPUTATIONAL RESULTS OF DIFFERENT SYSTEMS.

| System | Objective | Constraint violation | Solution time/s | Iteration | Gap/% |
|---|---|---|---|---|---|
| 9  | 0.0176 | 4.37e-11 | 0.3490 | 13  | 5.31e-08 |
| 14 | 0.1044 | 1.56e-10 | 0.5250 | 34  | 2.54e-07 |
| 30 | 0.0799 | 2.18e-11 | 1.3340 | 18  | 5.28e-07 |
| 57 | 0.0740 | 1.30e-12 | 10.537 | 291 | 6.69e-07 |

From Tab. 1, as the grid size increases, the number of variables and constraints of the model increases rapidly, and the solution time of the problem increases. However, even for IEEE 57-bus system, the solution time is still within 1.5 minutes, which is acceptable for industrial applications. The computation can be completed within each period (e.g., 5 min) to update the optimal references of the controllers and overcome the effects of power fluctuations in the IBRs.

## VI. CONCLUSION

In this paper, the dynamic voltage support (DVS) capability of IBRs under large disturbances is analyzed, and it is found that their weak DVS capability is caused by the lack of electromagnetic couplings in the windings of SMs. By modifying the inner loop of the controllers based on the SM transient model, the DVS effect is effectively improved. Due to the variety of inverter-based resources' (IBRs) control strategies and the dispersion of IBRs' locations, their voltage support effects on weak points may compete. By formulating a multi-IBRs control parameter optimization model, the capacity of different inverters can be effectively aggregated to improve the overall STVS of the system with the same capacity configuration. Inspired by singular perturbation theory, the fast-dynamic related differential algebraic equations (DAEs) in the model are degraded to algebraic equations, while the slow-dynamic related state variables remain constant. Eventually, the original model is transformed into a general non-linear programming (NLP) model and the Hessian matrix is constant. The advanced solver IPOPT is used for testing. Test results based on different systems show that the proposed method can find the optimal solution of the model efficiently, and the solution results can guarantee the STVS of the system under the anticipated faults. In the future, the coordinated control method of multi-IBRs considering various types of stability will be further investigated for 100% IBR penetrated power systems.